\documentclass[11pt]{article}

\usepackage[final]{acl}

\usepackage{times}
\usepackage{latexsym}
\usepackage[T1]{fontenc}
\usepackage[utf8]{inputenc}
\usepackage{microtype}
\usepackage{inconsolata}
\usepackage{graphicx}
\usepackage{booktabs}
\usepackage{amsmath,amssymb}

\newif\ifanonymous
\anonymousfalse
\ifanonymous
  \newcommand{\sys}{\textsc{Unite}}
  \newcommand{\sysv}{\textsc{Unite}-v1}
  \newcommand{\figdir}{figures_anon}
  
\else
  \newcommand{\sys}{UIBK-Embedding}
  \newcommand{\sysv}{UIBK-Embedding-v1}
  \newcommand{\figdir}{figures}
  
\fi
\newcommand{\maxsim}{\operatorname{MaxSim}}

\title{UIBK-Embedding: Effective Text Retrieval by Unifying Dense and Late-Interaction Retrieval
}

\ifanonymous
\author{Anonymous submission}
\else
\author{Abdelrahman Abdallah, Mohammed Ali,  Adam Jatowt \\
   University of Innsbruck \\
  \texttt{Abdelrahman.Abdallah@uibk.ac.at}}
\fi

\begin{document}
\maketitle


\begin{abstract}
Dense bi-encoders and late-interaction retrievers are usually treated as alternatives, yet they fail on different queries: one vector captures what a text is about, token-level matching finds its exact entities and phrases.
Using both today means two models, two indexes and a score-fusion step.
We present \sysv{}, a 179M-parameter English encoder that produces both representations in one pass and scores them with one unmodified MaxSim.
The dense embedding is written into the list of token vectors as three extra vectors, and a one-dimensional tag keeps the two kinds of vectors from matching each other, so that plain MaxSim equals the token-level MaxSim plus a weighted dense cosine.
The model therefore runs in a standard late-interaction index without changes; we test it with PLAID.
A shared 18-layer trunk with a separate 4-layer top per head, and a cross-encoder distilled into the \emph{fused} score, turn the two heads into one retriever.
On BEIR, \sysv{} reaches 59.79 nDCG@10. This is 2.57 points above LateOn, the strongest model of its size, and 2.10 points above it when both models read documents of the same length; \sysv{} is ahead on all 15 datasets.
On a decontaminated version of BEIR it stays ahead on average (61.55 against 60.36) and on 8 of 12 datasets.
Among the 156 models with complete BEIR results in the MTEB results repository, only models with at least 1.5B parameters score higher.
Code, model weights, and a technical blog post are available on \href{https://github.com/DataScienceUIBK/uibk-embedding}{GitHub}, \href{https://huggingface.co/DataScience-UIBK/uibk-embedding-v1}{Hugging Face}, and our \href{https://huggingface.co/blog/abdoelsayed/uibk-embedding-v1}{project blog}.
\end{abstract}

\begin{figure}[t]
\centering
\includegraphics[width=\columnwidth]{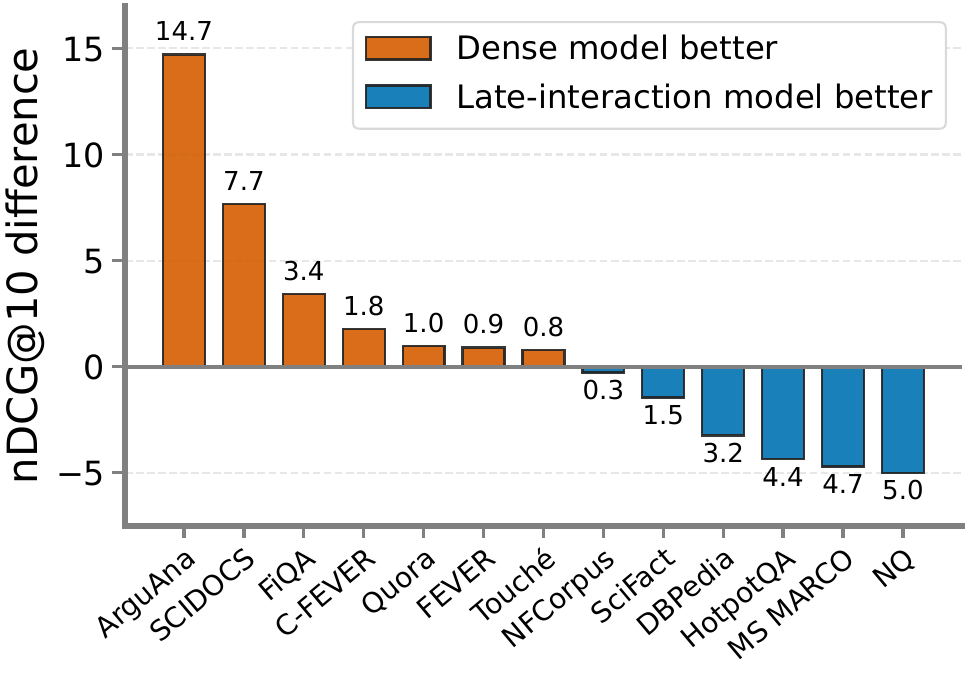}
\caption{A dense model and a late-interaction model, fine-tuned from the same backbone on the same data, compared on the 13 NanoBEIR datasets. Each bar is the difference of their nDCG@10. The two models win on different datasets.}
\label{fig:complementarity}
\end{figure}

\section{Introduction}
\label{sec:intro}

Neural retrievers encode documents offline and compare them with the query at search time, and two families of models dominate.
\emph{Dense bi-encoders} compress a text into a single vector and score a query--document pair with one cosine \citep{dpr,sbert,e5,gte,bge,abdallah2026rankify}.
\emph{Late-interaction} models keep one vector per token and score a pair with MaxSim, the sum over the query tokens of their best-matching document token \citep{colbert,colbertv2,plaid}.
The two are usually presented as a choice between a small index and better generalisation \citep{beir}.

The two families do not make the same mistakes. Figure~\ref{fig:complementarity} compares a dense model and a ColBERT model that we fine-tuned from the same backbone on the same data. The dense model is better where a long query has to be matched as a whole, as in argument retrieval, and the late-interaction model is better where a single entity or phrase decides, as in the open-domain question answering. Their scores are complementary. Systems exploit this complementarity today through \emph{hybrid} retrieval: two models, two indexes, two searches, and a fusion step such as reciprocal rank fusion \citep{rrf}. Models that return several representations at once \citep{bgem3,jinav4} save the second encoder, but every representation still has its own index and its own scoring function.

\begin{figure*}[t]
\centering
\includegraphics[width=\textwidth]{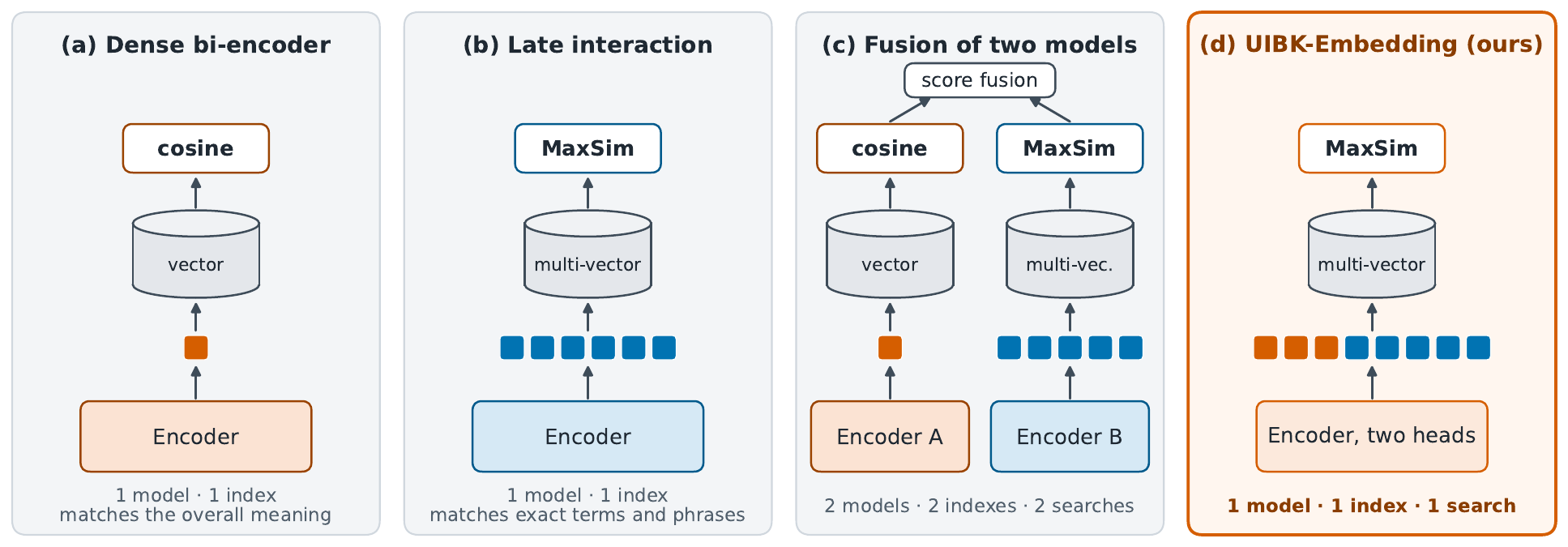}
\caption{Four ways to retrieve. (a) A dense bi-encoder stores one vector per text. (b) A late-interaction model stores one vector per token. (c) Hybrid retrieval runs both and fuses their scores. (d) \sys{} computes both scores with one encoder, one index and one MaxSim.}
\label{fig:paradigms}
\end{figure*}

In this work we introduce \textbf{\sys{}}, a retriever that holds a dense head and a late-interaction head in one encoder and adds their scores inside a standard late-interaction index (Figure~\ref{fig:paradigms}d).
Specifically, we make the following contributions: \textbf{(1)} For the representation, we store the dense sentence vector as three extra vectors, the \emph{slots}, in the list of token vectors, and give every vector a one-dimensional \emph{tag} with opposite signs for tokens and slots. Inside MaxSim a token then never selects a slot and a slot never selects a token, so the unmodified MaxSim returns the token-level score plus a weighted dense cosine (\S\ref{sec:onelist}). The index and the scorer need no change: the model is saved as a ColBERT model and runs in PLAID \citep{plaid} through PyLate \citep{pylate}.
\textbf{(2)} For the architecture, the two heads share the first 18 layers of ModernBERT-base \citep{modernbert} and each has its own copy of the last four, which raises the score of each head on its own by about one point. The dense head uses the latent-attention pooling of NV-Embed \citep{nvembed}, here on an encoder-only model (\S\ref{sec:encoder}). \textbf{(3)} For the training, we distil a cross-encoder into the \emph{fused} score, the score that is used at inference, in addition to the losses of each head (\S\ref{sec:training}). Ablations show the contribution of the two heads, of the fused distillation and of the main training choices (\S\ref{sec:analysis}).  \textbf{(4)} \sysv{} has 179M parameters and reaches 59.79 nDCG@10 on BEIR \citep{beir} and 70.55 on NanoBEIR \citep{nanobeir}, evaluated with the public MTEB implementation \citep{mteb}. It is 2.57 points above LateOn \citep{lateon}, the strongest model of its size, 2.10 points above it at equal document length, and ahead of it on all 15 datasets. On the decontaminated BEIR of \citet{lateon} it reaches 61.55, against 60.36 for LateOn, and is ahead on 8 of the 12 datasets. See Tables~\ref{tab:beir} and~\ref{tab:decon} for details.

\section{Related Work}
\label{sec:related}

\paragraph{Dense retrieval.}
Bi-encoders trained with contrastive learning \citep{dpr,sbert,contriever} improved through hard-negative mining \citep{ance,nvretriever}, large weakly supervised pre-training \citep{e5,gte,bge,nomic,arctic} and distillation from cross-encoders \citep{crossarch,rocketqav2}.
NV-Embed \citep{nvembed} replaces mean or [CLS] pooling with a latent-attention layer in the spirit of the Perceiver \citep{perceiver}; our dense head uses this pooling on an encoder-only model.

\paragraph{Late interaction.}
ColBERT \citep{colbert} introduced MaxSim over token vectors; ColBERTv2 \citep{colbertv2} added residual compression and cross-encoder distillation, and PLAID \citep{plaid} made the search fast.
Later models extend it to long contexts and many languages \citep{jinacolbertv2,jacolbert,xtr}.
On ModernBERT \citep{modernbert}, SmallReason-ColBERT~\cite{abdallah2026smallreason}, GTE-ModernColBERT \citep{gtemoderncolbert}, EXCISE~\cite{ali2026excise},ColBERT-Zero \citep{colbertzero} and LateOn \citep{lateon} are the strongest models of their size, and LateOn is our reference.

\paragraph{Hybrid and unified retrievers.}
Combining lexical and dense signals is standard practice \citep{mebert,splade,rrf}.
COIL \citep{coil} adds a [CLS] similarity to contextualised exact-match scores, stored in inverted lists.
ColBERTer \citep{colberter} retrieves with a single vector and refines with token vectors.
UnifieR \citep{unifier} learns dense and lexical views together, and BGE-M3 \citep{bgem3} and jina-embeddings-v4 \citep{jinav4} return dense and multi-vector outputs from one model.
All of these keep separate scoring functions and index structures, or use the signals in separate stages.
\sys{} differs in that both signals live in one list of vectors and are added by the standard MaxSim, so one existing index and one search serve both.
For single-vector retrieval, \citet{mebert} observe that a weighted sum of a sparse and a dense score is one inner product of concatenated vectors.
MaxSim is a sum of maxima over many vectors, where concatenation does not apply; our contribution is the construction that makes the same additive hybrid hold there.
BGE-M3 distils the sum of its own scores into each of its heads; we go the other way and distil an external cross-encoder into the sum.

\section{Background}
\label{sec:background}

Let $q$ be a query and $d$ a document. All vectors below have unit norm.

\paragraph{Dense retrieval.} An encoder maps every text to one vector, $e_q$ and $e_d$, and the score is their cosine,
\begin{equation}
s_{\text{den}}(q,d) = e_q^{\top} e_d .
\end{equation}

\paragraph{Late interaction.} The encoder returns one vector per token, $u_1,\dots,u_m$ for the query and $u'_1,\dots,u'_n$ for the document. The score matches every query token with its most similar document token \citep{colbert}:
\begin{equation}
s_{\text{tok}}(q,d) = \sum_{i=1}^{m} \max_{1 \le j \le n} u_i^{\top} u'_j .
\label{eq:late}
\end{equation}
A multi-vector index such as PLAID \citep{plaid} stores the token vectors of all documents and evaluates Eq.~\ref{eq:late}.

\paragraph{Hybrid retrieval.} A hybrid system computes both scores and combines them, for example as $s_{\text{tok}} + \lambda\, s_{\text{den}}$ or by fusing the two rankings \citep{rrf}. Both representations must then exist at search time, which means two encoders, or one encoder with two outputs, and in either case two indexes that are searched separately (Table~\ref{tab:paradigms}).

\begin{table}[t]
\centering
\small
\setlength{\tabcolsep}{3pt}
\resizebox{\columnwidth}{!}{%
\begin{tabular}{lccccc}
\toprule
 & Enc. & Indexes & Searches & Sentence & Token \\
\midrule
Dense bi-encoder & 1 & 1 & 1 & \checkmark & -- \\
Late interaction & 1 & 1 & 1 & -- & \checkmark \\
Fusion of two models & 2 & 2 & 2 & \checkmark & \checkmark \\
Multi-output encoder & 1 & 2 & 2 & \checkmark & \checkmark \\
\textbf{\sys{} (ours)} & 1 & 1 & 1 & \checkmark & \checkmark \\
\bottomrule
\end{tabular}}
\caption{What a retriever needs at search time, and which signals it scores: a sentence-level match, a token-level match, or both. A multi-output encoder returns a dense vector and token vectors from one model \citep{bgem3,jinav4}.}
\label{tab:paradigms}
\end{table}

Our goal is to compute $s_{\text{tok}} + \lambda\, s_{\text{den}}$ with one encoder and with the machinery of late interaction alone: one multi-vector index, one search, and Eq.~\ref{eq:late} as the only scoring function.

\section{\sys{}}
\label{sec:method}

\begin{figure*}[t]
\centering
\includegraphics[width=\textwidth]{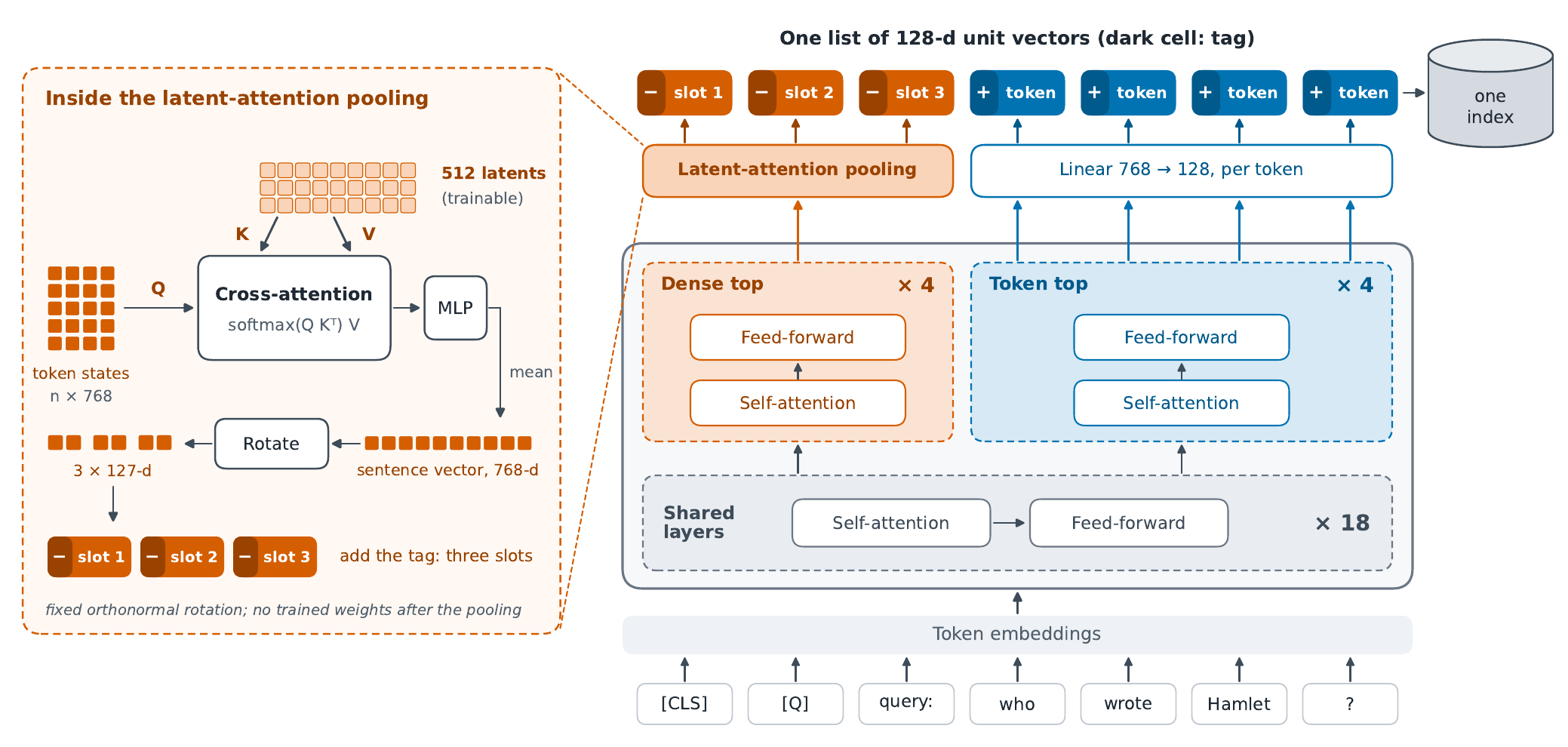}
\caption{\sysv{} is one encoder (right, drawn bottom-up). The first 18 layers are shared; the last four layers exist twice, one copy for each head. The token head projects every token to 128 dimensions. The dense head is a latent-attention pooling (left): the token states of the dense top attend to 512 trainable latents, and the mean of the result is the sentence vector, which a fixed rotation splits into three \emph{slots}. The slots take the first three positions of the output, so the model returns one list of 128-dimensional unit vectors whose first coordinate is a tag ($-$ for slots, $+$ for tokens), stored in one standard late-interaction index.}
\label{fig:architecture}
\end{figure*}

Figure~\ref{fig:architecture} shows the model: a text is encoded once, and the output is an ordinary list of 128-dimensional unit vectors.

\subsection{One Trunk, Two Tops}
\label{sec:encoder}

The backbone is ModernBERT-base \citep{modernbert} with 22 layers.
The embeddings and the first 18 layers form a trunk that both heads share.
Each head then has its own copy of the last four layers: the \emph{token top} is the original one, the \emph{dense top} is a copy that is trained separately.
For a text with tokens $x_1,\dots,x_n$ the encoder returns two sequences of states,
\begin{equation}
H^{\text{tok}} = T_{\text{tok}}(E(x)), \qquad H^{\text{den}} = T_{\text{den}}(E(x)),
\end{equation}
where $E$ is the trunk and $T_{\text{tok}}$, $T_{\text{den}}$ are the two tops.
The second top adds 20.1M parameters to the 149.0M of the backbone, and the cost of one pass grows from 22 to 26 layer evaluations.

\paragraph{Token head.} As in ColBERT, a linear layer maps every state to 128 dimensions, $\hat u_t = W h^{\text{tok}}_t / \lVert W h^{\text{tok}}_t\rVert$.

\paragraph{Dense head.} A latent-attention layer \citep{nvembed} lets every state of the dense top attend to 512 trainable latent vectors, followed by a feed-forward layer; the mean of the result over the tokens is the sentence vector $g \in \mathbb{R}^{768}$.
In a controlled comparison with the same data, steps and seed, in which the two training-loss curves coincide, this pooling was 0.8 NanoBEIR points better than [CLS] pooling.

\subsection{One List of Vectors, One MaxSim}
\label{sec:onelist}

A late-interaction index stores 128-dimensional vectors, and its only scoring function is
\begin{equation}
\maxsim(q,d) = \sum_{i \in q} \max_{j \in d}\; v_i^{\top} v'_j .
\label{eq:maxsim}
\end{equation}
We want this function to return a token-level score plus a dense score. Two things are in the way: the sentence vector has 768 dimensions, and a query token must not be matched with the sentence vector of a document.

\paragraph{Slots.} A fixed random matrix $R \in \mathbb{R}^{381 \times 768}$ with orthonormal rows projects $g$ to $3 \times 127$ dimensions, which are cut into three slices and normalised: $g_s = \operatorname{norm}\big((Rg)_{[127(s-1)+1\,:\,127s]}\big)$, $s=1,2,3$.
The sum of the three slice cosines of two texts approximates the cosine of their sentence vectors.
The token vectors are reduced to 127 dimensions by a fixed matrix $R' \in \mathbb{R}^{127 \times 128}$ with orthonormal rows and normalised again, $u_t = \operatorname{norm}(R'\hat u_t)$.

\paragraph{Tag.} Every vector gets one more coordinate, placed first:
\begin{equation}
\begin{aligned}
\text{token } t:&\quad v_t = \big[\, {+\sqrt{1-\beta}}\;;\; \sqrt{\beta}\,u_t \,\big],\\
\text{slot } s:&\quad v_s = \big[\, {-\sqrt{1-\delta}}\;;\; \sqrt{\delta}\,g_s \,\big].
\end{aligned}
\end{equation}
All vectors have unit norm. The inner products are
\begin{equation}
\begin{aligned}
v_t^{\top} v'_{t'} &= (1-\beta) + \beta\, u_t^{\top}u'_{t'},\\
v_s^{\top} v'_{s} &= (1-\delta) + \delta\, g_s^{\top}g'_{s},\\
v_t^{\top} v'_{s} &= -\sqrt{(1-\beta)(1-\delta)} + \sqrt{\beta\delta}\, u_t^{\top}g'_{s}.
\end{aligned}
\label{eq:products}
\end{equation}
We use $\beta = 0.9$ and $\delta = 0.225$. A pair of the same kind then gets a positive constant ($0.1$ for tokens, $0.775$ for slots) and a mixed pair a negative one ($-0.28$). Three properties follow from Eq.~\ref{eq:products}.

\emph{(i) A slot never selects a token.} A slot--slot product is at least $(1-\delta)-\delta = 0.55$, and a slot--token product is at most $-\sqrt{(1-\beta)(1-\delta)}+\sqrt{\beta\delta} = 0.17$. This holds for every input.

\emph{(ii) A token never selects a slot} if its best token match has a cosine above $0.08$: the token--token product is then above $0.1 + 0.9 \cdot 0.08 = 0.17$, the largest value a token--slot product can take. This bound assumes the worst case $u^{\top}g = 1$. The two vectors come from unrelated 127-dimensional spaces, where a typical product has magnitude $1/\sqrt{127} \approx 0.09$; for $|u^{\top}g| \le 0.3$ the condition relaxes to a best token cosine above $-0.27$.

\emph{(iii) A slot selects its own slot} whenever $g_s^{\top}g'_s \ge g_s^{\top}g'_{s'}$ for the other two slots $s'$. These hold different coordinates of the sentence vector, so the right-hand side is near zero. The condition fails only when the sentence vectors of the two texts are unrelated, and the dense term is then floored near zero instead of turning negative.

In practice the tag separates the two kinds completely: on NanoBEIR, over all pairs of a query vector and a document, no query token selects a slot and no slot selects a token.
The three properties are statements about exact inner products. We use the model with PLAID, which selects candidates from compressed vectors and re-scores them with MaxSim, and the benchmark results in this paper are obtained that way; we make no claim for indexes that transform the vectors in other ways.

\paragraph{The score.} Inserting Eq.~\ref{eq:products} into Eq.~\ref{eq:maxsim} gives
\begin{equation}
\begin{aligned}
\maxsim(q,d) \;=\;& C(q) + \beta\,\Big(\underbrace{\sum_{t \in q}\max_{t' \in d} u_t^{\top}u'_{t'}}_{\text{late interaction}}\\
&+\; \lambda \cdot \underbrace{\tfrac{1}{3}\sum_{s=1}^{3} g_s^{\top}g'_s}_{\text{dense}}\Big),
\end{aligned}
\label{eq:fused}
\end{equation}
with $\lambda = 3\delta/\beta = 0.75$ (Figure~\ref{fig:scoring} in the appendix). The constant $C(q) = (1-\beta)\,m + 3(1-\delta)$, for a query of $m$ tokens, does not depend on the document, so the ranking is that of the term in parentheses.
The weight of the dense score is thus a property of the stored vectors and can be chosen when the model is packaged, without training.

\paragraph{Practical points.} The slots are written over the first three positions of every text, which hold the [CLS] token, the query or document marker and the prompt word, so the list is not longer than that of a ColBERT model.
The dense head reads queries of up to 256 tokens, while only the first 64 query tokens take part in the token-level matching.
Documents are read up to 2048 tokens.



\subsection{Training}
\label{sec:training}

\paragraph{Stages.} (1) \emph{Pre-training.} We pre-train ModernBERT-base with the latent-attention pooling contrastively for 33k steps with a batch of 16{,}384 pairs on a 331M-pair sample of the public pre-training collection of \citet{lateon}.
(2) \emph{Single-head models.} From this backbone we fine-tune a dense model (the weight average \citep{soups} of four runs) and a ColBERT model; they are the two models of Figure~\ref{fig:complementarity}.
(3) \emph{\sys{}.} The two encoders of stage 2 are fine-tunes of one backbone, so a hybrid model can be initialised from their weight average. It is trained with both heads for three epochs; its last four layers are then duplicated into the two tops, and the two-top model is trained for one final epoch.
\begin{figure}[t]
\centering
\includegraphics[width=\columnwidth]{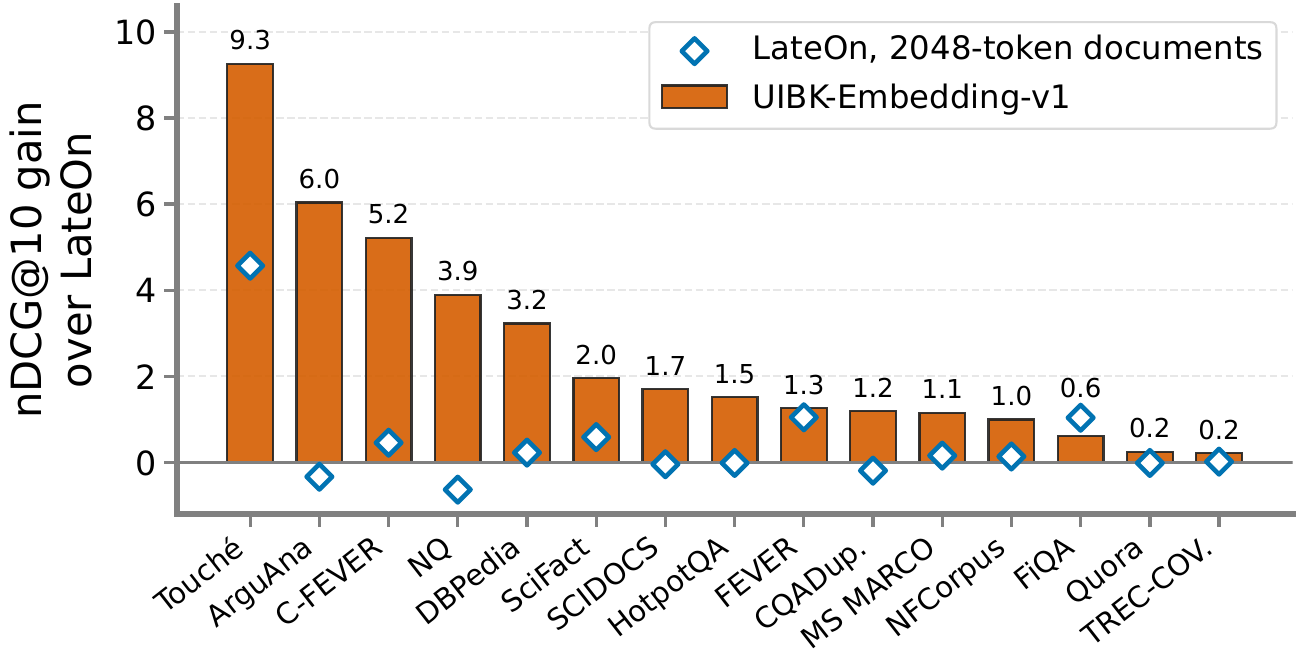}
\caption{Gain over LateOn on every BEIR dataset. Diamonds: the gain LateOn itself obtains when it reads documents of up to 2048 tokens instead of 300.}
\label{fig:perdataset}
\end{figure}
\paragraph{Loss.} For a query with its positive and 14 mined hard negatives, let $s^{\text{tok}}$, $s^{\text{den}}$ and $s^{\text{fus}}$ be the token-level, dense and fused (Eq.~\ref{eq:fused}) scores, and $y$ the scores of a cross-encoder teacher \citep{mxbairerank}. The loss is
\begin{equation}
\begin{aligned}
\mathcal{L} = {}& \mathcal{L}_{\text{NCE}}(s^{\text{tok}}) + \mathcal{L}_{\text{NCE}}(s^{\text{den}})\\
&+ \mathrm{KL}\big(y \,\Vert\, s^{\text{den}}\big) + \mathrm{KL}\big(y \,\Vert\, s^{\text{fus}}\big) + \gamma\,\mathcal{L}_{\text{anchor}} .
\end{aligned}
\label{eq:loss}
\end{equation}
$\mathcal{L}_{\text{NCE}}$ is the contrastive loss over all documents of the batch (temperature 0.02).
The KL terms compare the softmax of the student scores over the candidates of the query with that of the teacher \citep{hinton,crossarch}.
$\mathcal{L}_{\text{anchor}}$ is one minus the cosine between the vectors of each head and those of the single-head model of stage 2, for queries and documents ($\gamma = 10$).
The dense score $s^{\text{den}}$ is the mean cosine of the three slots, the dense term of Eq.~\ref{eq:fused}, so training and inference use the same quantity.
The term $\mathrm{KL}(y \Vert s^{\text{fus}})$ is computed on the score that is used at inference; it is the only term in which the two heads appear together.

\begin{table*}[t]
\centering
\resizebox{\textwidth}{!}{%
\begin{tabular}{lrccccccccccccccc|c}
\toprule
Model & Size & Arg. & CQA & C-FEV & DBP & FEV & FiQA & HQA & MSM & NFC & NQ & Quora & SCID & SciF & COV & Tou. & Avg. \\
\midrule
\multicolumn{18}{l}{\textit{Single-vector models, at most 250M parameters}} \\
bge-base-en & 109M & 61.49 & 41.35 & 31.61 & 42.01 & 88.35 & 40.76 & 72.35 & 41.26 & 35.54 & 53.98 & 88.78 & 21.30 & 73.26 & 76.45 & 26.54 & 53.00 \\
modernbert-embed-base & 149M & 48.96 & 42.08 & 35.67 & 41.49 & 87.35 & 40.59 & 67.11 & 41.47 & 33.40 & 62.15 & 88.85 & 18.59 & 69.63 & \underline{84.15} & 31.91 & 52.89 \\
granite-embedding-english-r2 & 149M & 59.21 & \textbf{50.00} & 35.82 & 39.60 & 88.04 & 46.32 & 67.36 & 32.14 & 37.49 & 58.22 & 87.84 & \textbf{24.95} & 75.80 & 70.56 & 22.90 & 53.08 \\
gte-base-en-v1.5 & 137M & 63.49 & 39.52 & 40.36 & 39.90 & \textbf{94.81} & 48.65 & 67.75 & 42.62 & 35.88 & 52.96 & 88.42 & 21.92 & \underline{76.77} & 73.13 & 25.22 & 54.09 \\
cadet-embed-base-v1 & 109M & 56.16 & 40.80 & 36.69 & 45.26 & 89.33 & 40.89 & 73.76 & 42.56 & 37.71 & 59.32 & 87.59 & 19.09 & 74.68 & 80.55 & 30.42 & 54.32 \\
snowflake-arctic-embed-m & 109M & 56.44 & 43.81 & 39.37 & 44.73 & 89.02 & 42.40 & 73.65 & 41.77 & 36.77 & 62.43 & 87.42 & 21.10 & 73.55 & 79.65 & 31.47 & 54.91 \\
gte-modernbert-base & 149M & \textbf{74.56} & 42.64 & \textbf{45.90} & 41.39 & \underline{93.97} & 49.54 & 70.39 & 39.93 & 34.32 & 56.10 & 88.56 & 20.44 & 76.41 & 75.75 & 17.97 & 55.19 \\
jina-embeddings-v5-text-nano & 212M & \underline{65.70} & 44.65 & 39.60 & 45.26 & 89.51 & 47.85 & 69.07 & 41.64 & 38.69 & 63.38 & 88.87 & 22.60 & 75.78 & 77.60 & 30.70 & 56.06 \\
DenseOn & 149M & 54.65 & 46.89 & 37.49 & 44.65 & 90.69 & \textbf{53.86} & 74.51 & 43.58 & \textbf{39.03} & 59.25 & 89.31 & 22.35 & 75.95 & 82.33 & 28.43 & 56.20 \\
\midrule
\multicolumn{18}{l}{\textit{Late-interaction models, at most 250M parameters}} \\
GTE-ModernColBERT-v1 & 149M & 47.52 & 41.08 & 31.33 & \underline{47.56} & 87.67 & 45.25 & 77.48 & 45.60 & 37.83 & 61.62 & 86.71 & 19.22 & 76.33 & \textbf{84.84} & 31.25 & 54.75 \\
ColBERT-Zero & 149M & 52.81 & 41.41 & 35.90 & 47.43 & 90.52 & 42.50 & 79.45 & \underline{45.95} & 37.21 & 61.82 & 85.19 & 19.84 & 76.33 & 78.27 & \underline{36.24} & 55.39 \\
LateOn & 149M & 50.52 & 47.36 & 39.67 & 45.99 & 92.02 & 53.12 & \underline{79.98} & 45.67 & 37.79 & \underline{63.91} & \underline{89.66} & 21.90 & 76.61 & 83.60 & 30.52 & \underline{57.22} \\
\textbf{UIBK-Embedding-v1 (ours)} & 179M & 56.56 & \underline{48.55} & \underline{44.89} & \textbf{49.22} & 93.28 & \underline{53.73} & \textbf{81.50} & \textbf{46.82} & \underline{38.79} & \textbf{67.80} & \textbf{89.92} & \underline{23.60} & \textbf{78.57} & 83.82 & \textbf{39.78} & \textbf{59.79} \\
\bottomrule
\end{tabular}

}
\caption{BEIR, nDCG@10 on every dataset, for the models with at most 250M parameters. All numbers are from the official MTEB results repository; ours are produced by the same MTEB code. \textbf{Bold}: best; \underline{underlined}: second best. For the full table with larger models, see Table~\ref{tab:beir-full} in Appendix~\ref{app:results}.}
\label{tab:beir}
\end{table*}

\subsection{Training Data}
\label{sec:data}

\paragraph{Supervised collection.} We use the English fine-tuning collection of \citet{lateon}: 1.56M queries from MS~MARCO \citep{msmarco}, NQ, HotpotQA, FEVER, FiQA, SQuAD~v2, TriviaQA and MIRACL, each with mined negatives and the scores of the cross-encoder.

\paragraph{Re-mined negatives.} Hard negatives are most useful when they are hard for the model that is trained. For every query we retrieve five more candidates with our own dense model, score them with the cross-encoder and keep the four hardest whose score is at least one logit below that of the positive. This adds 3.59M negatives, and every query then has 14 negatives with teacher scores.

\paragraph{Auxiliary pairs.} The supervised collection consists of questions over Wikipedia and web passages. To cover duplicate questions, community forums and scientific literature as well, we mine 412k pairs from the pre-training collection: Quora duplicate questions (35k), StackExchange duplicates, question--answer and title--body pairs (246k), and S2ORC citation and title--abstract pairs (131k). Negatives come from our dense model with the positive-aware filter of \citet{nvretriever}: candidates that score above 95\% of the positive are dropped. These sources are in-domain for Quora, CQADupStack, SCIDOCS and SciFact: the BEIR test queries are removed, but the documents come from the same sites.

\paragraph{Long queries.} The supervised queries are short questions, but a query can also be a whole paragraph, as in argument retrieval. We add 152k pairs whose queries are long, also from the pre-training collection: Reddit posts with a reply (39k), S2ORC abstracts with the abstract of a cited paper (57k) and StackExchange question bodies with a related body (56k). None of them comes from the sources of ArguAna or Touch\'e. We train with queries of up to 256 tokens, which the dense head reads in full.


\section{Experimental Setup}
\label{sec:setup}

\paragraph{Benchmarks.} We evaluate on the 15 datasets of BEIR \citep{beir} and on NanoBEIR \citep{nanobeir}, which consists of 50-query samples of 13 BEIR datasets, and report nDCG@10.
All BEIR and NanoBEIR numbers of \sysv{} are produced by MTEB \citep{mteb} (versions 2.21.6 and 2.24.8, which give the same scores) with its standard late-interaction wrapper and a PLAID index with default settings; no part of the evaluation is specific to our model.
Queries are encoded up to 256 tokens and documents up to 2048, in bfloat16.

\paragraph{Baselines.} The numbers of all other models are taken from the official MTEB results repository \citep{mtebresults}, which holds complete BEIR results for 156 models and complete NanoBEIR results for 20.
We compare in detail with the models of the same size class, at most 250M parameters, and list larger models in Appendix~\ref{app:results}.

\paragraph{Training.} The last stage takes 4.3 hours on four H100 GPUs and the pre-training of stage 1 about 75 hours; Appendix~\ref{app:details} lists all settings.

\section{Results}
\label{sec:results}

\begin{table*}[t]
\centering
\resizebox{\textwidth}{!}{%
\begin{tabular}{lrcccccccccccc|c}
\toprule
Model & Size & Arg. & DBP & FiQA & HQA & MSM & NFC & NQ & Quora & SCID & SciF & COV & Tou. & Avg. \\
\midrule
answerai-colbert-small & 33M & 47.63 & 30.38 & 45.75 & 76.35 & 72.39 & 25.41 & 86.10 & 89.36 & 12.72 & 87.82 & 77.51 & 27.81 & 56.60 \\
ColBERT-Zero & 149M & 54.49 & \textbf{32.95} & 46.55 & 77.84 & \textbf{74.21} & 26.59 & 91.13 & 88.29 & 14.22 & 89.48 & 75.32 & 40.93 & 59.33 \\
LateOn (published) & 149M & 52.16 & 31.72 & \textbf{57.92} & 78.94 & 70.31 & 26.97 & \textbf{93.05} & 91.54 & 15.09 & 88.92 & 80.94 & 36.77 & 60.36 \\
Qwen3-Embedding-0.6B & 596M & 48.40 & 25.30 & 49.10 & 62.20 & 63.60 & 25.80 & 88.30 & 90.00 & 15.30 & 85.50 & \textbf{87.90} & 31.80 & 56.10 \\
jina-embeddings-v5-text-nano & 212M & 47.22 & 30.18 & 51.45 & 67.48 & 68.58 & \textbf{29.38} & 92.26 & 91.33 & 14.93 & 89.36 & 76.80 & 33.16 & 57.68 \\
DenseOn & 149M & 40.02 & 28.79 & 55.87 & 73.65 & 68.91 & 28.54 & 92.13 & 91.09 & 14.71 & 85.38 & 82.45 & 31.03 & 57.71 \\
pplx-embed-v1-0.6b & 596M & 43.66 & 28.44 & 55.17 & 73.50 & 71.87 & 27.99 & 91.63 & 91.54 & 15.44 & 89.02 & 83.68 & 30.05 & 58.50 \\
\midrule
LateOn (our run) & 149M & 46.90 & 31.42 & 57.71 & 79.11 & 69.72 & 27.17 & 91.63 & 91.53 & 15.09 & \textbf{89.92} & 81.00 & 35.06 & 59.69 \\
\textbf{UIBK-Embedding-v1 (ours)} & 179M & \textbf{56.51} & 31.39 & 56.19 & \textbf{80.32} & 71.74 & 29.32 & 90.45 & \textbf{91.81} & \textbf{16.91} & 87.78 & 81.54 & \textbf{44.61} & \textbf{61.55} \\
\bottomrule
\end{tabular}

}
\caption{Decontaminated BEIR \citep{lateon}, nDCG@10 on its 12 datasets. The numbers of the other models are those published with the benchmark; the last two rows are evaluated by us under one protocol. \textbf{Bold}: best per column.}
\label{tab:decon}
\end{table*}

\paragraph{BEIR.} Table~\ref{tab:beir} gives the results on every dataset.
\sysv{} reaches 59.79, against 57.22 for LateOn, the best model of its size class, 56.20 for the dense DenseOn and 56.06 for jina-embeddings-v5-text-nano.
Among the 13 models of the table it is best on 7 of the 15 datasets and second on 5 more.
It is also ahead of the larger models of up to 600M parameters, such as mLateOn (57.56, 307M), jina-embeddings-v5-text-small (56.67, 596M) and Qwen3-Embedding-0.6B (55.52); Table~\ref{tab:beir-full} in the appendix lists them.
Among all 156 models with complete BEIR results it ranks tenth, and the nine models above it have between 1.5B and 7.6B parameters, at least eight times more.
On the ten retrieval tasks of MTEB(eng, v2), which are derived from BEIR, it reaches 62.47, the best of the 45 models with at most 250M parameters and 13th of the 134 models with all ten tasks.

\paragraph{Every dataset improves.} \sysv{} is ahead of LateOn on all 15 datasets (Figure~\ref{fig:perdataset}).
The margins range from 0.2 points on TREC-COVID and Quora to 9.3 on Touch\'e; the next largest gains are on ArguAna ($+6.0$), Climate-FEVER ($+5.2$), NQ ($+3.9$) and DBPedia ($+3.2$).

\paragraph{Not an effect of document length.} LateOn's reported numbers use documents of up to 300 tokens.
With 2048 tokens LateOn reaches 57.69 in our evaluation (diamonds in Figure~\ref{fig:perdataset}), so the gap at equal document length is 2.10 points.

\paragraph{Decontaminated BEIR.} To verify that the result does not rest on overlap between training data and the benchmark, we evaluate on the decontaminated BEIR of \citet{lateon}: 12 BEIR datasets from which the documents and queries that overlap with common training collections were removed.
We evaluate with PyLate and a PLAID index, as MTEB does, and run LateOn under the same protocol (Table~\ref{tab:decon}).
\sysv{} reaches 61.55. LateOn reaches 60.36 as published and 59.69 in our run, a difference that comes almost entirely from ArguAna, so the gain is between 1.2 and 1.9 points, and \sysv{} is ahead on 8 of the 12 datasets.

\begin{table*}[t]
\centering
\resizebox{\textwidth}{!}{%
\begin{tabular}{lrccccccccccccc|c}
\toprule
Model & Size & Arg. & C-FEV & DBP & FEV & FiQA & HQA & MSM & NFC & NQ & Quora & SCID & SciF & Tou. & Avg. \\
\midrule
BM25 & -- & 48.04 & 31.77 & 61.38 & 80.94 & 43.74 & \underline{82.77} & 53.30 & 32.50 & 50.07 & 86.28 & 33.51 & 70.99 & \textbf{69.53} & 57.29 \\
multilingual-e5-small & 118M & 44.54 & 30.64 & 60.54 & 83.49 & 43.53 & 81.01 & \underline{62.09} & 28.82 & 68.62 & \textbf{97.28} & 34.38 & 72.46 & 49.54 & 58.23 \\
F2LLM-v2-80M & 80M & 53.45 & 39.19 & 56.99 & \underline{92.89} & 44.89 & 72.66 & 56.91 & 32.63 & 54.59 & 95.86 & 38.54 & 66.89 & 53.84 & 58.41 \\
F2LLM-v2-160M & 159M & 57.07 & \underline{41.45} & 57.75 & 92.13 & 48.74 & 73.56 & 57.59 & 33.50 & 60.33 & 95.57 & \underline{40.59} & 67.76 & 53.11 & 59.93 \\
bekko-embedding-v1-a8m & 106M & 54.72 & 31.18 & 60.34 & 87.38 & 51.48 & 78.52 & 58.50 & 34.13 & 63.17 & 93.51 & 39.53 & 76.67 & 55.92 & 60.39 \\
bekko-embedding-v1-a25m & 123M & \textbf{60.73} & 26.06 & \underline{63.17} & 85.19 & \underline{53.39} & 80.91 & 61.46 & \underline{34.46} & \underline{69.10} & 94.24 & 39.46 & \underline{78.17} & 57.04 & \underline{61.80} \\
\textbf{UIBK-Embedding-v1 (ours)} & 179M & \underline{59.84} & \textbf{51.65} & \textbf{69.20} & \textbf{96.57} & \textbf{64.34} & \textbf{92.72} & \textbf{68.99} & \textbf{39.22} & \textbf{82.14} & \underline{96.99} & \textbf{45.40} & \textbf{83.93} & \underline{66.14} & \textbf{70.55} \\
\bottomrule
\end{tabular}

}
\caption{NanoBEIR, nDCG@10 on every dataset, for models with at most 250M parameters and official NanoBEIR results in the MTEB results repository. \textbf{Bold} and \underline{underlined} as in Table~\ref{tab:beir}. For the full table with larger models, see Table~\ref{tab:nanobeir-full} in Appendix~\ref{app:results}.}
\label{tab:nanobeir}
\end{table*}

\paragraph{NanoBEIR.} On NanoBEIR (Table~\ref{tab:nanobeir}) \sysv{} reaches 70.55, second among the 20 models with official results, behind inf-retriever-v1 (72.04, 7.1B parameters; Table~\ref{tab:nanobeir-full} in the appendix). Among the models with at most 250M parameters it is first on 10 of the 13 datasets and 8.75 points above the next model. LateOn has no official NanoBEIR results; with exact MaxSim search it scores 67.50 in our evaluation, against 70.43 for \sysv{}.

\paragraph{LoTTE.} On the test split of LoTTE \citep{colbertv2}, \sysv{} reaches a success@5 of 83.77, the highest among the models with official MTEB results (Table~\ref{tab:lotte}): 1.92 points above DenseOn-unsupervised and 16.51 above ColBERTv2. LoTTE is drawn from StackExchange, which is part of the public pre-training collection used by our model and by the two LightOn models; Appendix~\ref{app:results} gives the result per domain.

\begin{figure*}[th]
\centering
\includegraphics[width=\textwidth]{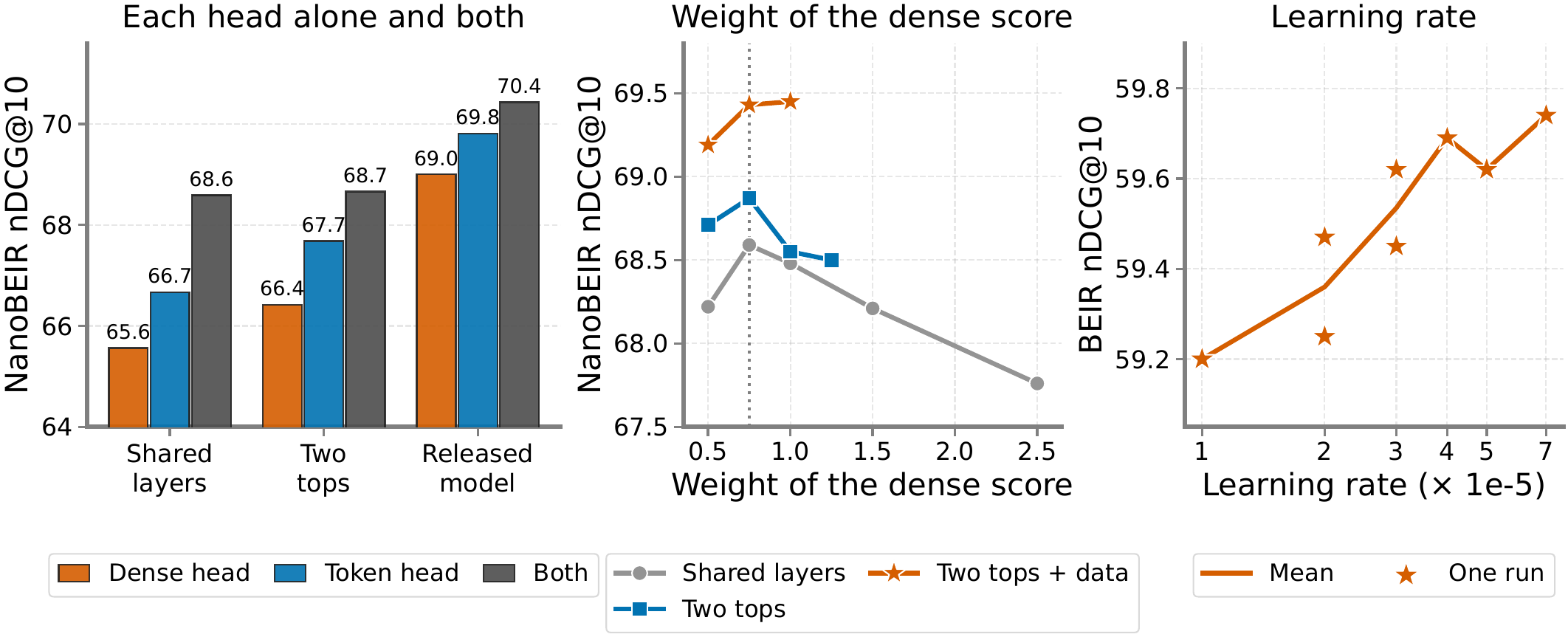}
\caption{Ablations. \textbf{Left:} each head alone and both together, for one encoder whose heads share all layers, one encoder with two tops, and the released model (NanoBEIR, exact MaxSim). \textbf{Middle:} the weight $\lambda$ of the dense score; the dotted line marks the released value. \textbf{Right:} learning rate of the last stage; stars are single runs.}
\label{fig:ablations}
\end{figure*}

\begin{table}[t]
\centering
\small
\begin{tabular}{lrc}
\toprule
Model & Size & Success@5 \\
\midrule
\textbf{\sysv{} (ours)} & 179M & \textbf{83.77} \\
DenseOn-unsupervised & 149M & 81.85 \\
ColBERT-Zero-unsupervised & 149M & 75.87 \\
gtr-t5-base & 110M & 69.84 \\
ColBERTv2 & 110M & 67.26 \\
BM25 & -- & 52.93 \\
\bottomrule
\end{tabular}
\caption{LoTTE, test split: success@5 averaged over the five domains and the two query types, for the models with official MTEB results.}
\label{tab:lotte}
\end{table}

\paragraph{LIMIT.} LIMIT \citep{limit} was built to show what a single vector cannot represent. On LIMIT small, \sysv{} reaches a recall@2 of 91.4, above GTE-ModernColBERT (83.5) and 37 points above the best single-vector model, which has 8B parameters (Table~\ref{tab:limit} in the appendix).

\paragraph{Cost.} We measure cost on FiQA (57{,}638 documents, one H100) with the same input lengths for every model, queries of 32 and documents of 300 tokens (Table~\ref{tab:cost} in the appendix). \sysv{} encodes 899 documents per second, against 986 for LateOn and 1{,}201 for DenseOn. One pass of our encoder is thus 1.7 times faster than the two passes of a LateOn and DenseOn hybrid (541 documents per second). Its index has the size of LateOn's (580 against 574~MiB) and is 22\% smaller than the two indexes of the hybrid (743~MiB). Search takes 7.8~ms per query in PLAID, against 4.4~ms for LateOn.

\section{Ablation Study}
\label{sec:analysis}

\paragraph{Both heads are needed.} In every setting of Figure~\ref{fig:ablations} (left) the fused score is 0.6 to 1.9 points above the better head.
The two tops raise each head on its own by about one point: 65.56 and 66.67 with shared layers, 66.42 and 67.68 with two tops.
For the released model, exact MaxSim gives 69.00 for the dense head alone, 69.81 for the token head alone and 70.43 for both.
The fused score has the best mean because each head alone falls far behind on some datasets: the token head reaches 51.66 on ArguAna against 67.39 for the dense head, and the dense head 34.26 on NFCorpus against 40.38 for the token head.

\paragraph{Weight of the dense score.} We vary $\lambda$ through $\delta$, with $\beta = 0.9$ fixed. The fused score is flat between $\lambda=0.75$ and $\lambda=1$ for every checkpoint we tested (Figure~\ref{fig:ablations}, middle), so we use $\lambda=0.75$. Appendix~\ref{app:bounds} gives the bounds for other weights.

\paragraph{Distillation into the fused score.} Adding the term $\mathrm{KL}(y \Vert s^{\text{fus}})$ to the losses of the two heads, with the data and the schedule unchanged, raises the BEIR average from 58.61 to 59.08. The gain is largest on Climate-FEVER ($+1.6$), ArguAna ($+1.1$) and NQ ($+1.0$), and with this term the model is ahead of LateOn on all 15 datasets.

\paragraph{Learning rate and seeds.} The learning rate of the last stage matters: $7\times10^{-5}$ is 0.54 points above $10^{-5}$ (Figure~\ref{fig:ablations}, right).
Two seeds of one configuration differ by about 0.2 points (59.25 and 59.47; 59.62 and 59.45), which is the resolution of the comparisons in this section.

\section{Conclusion}

We presented \sysv{}, a retriever that holds a dense head and a late-interaction head in one encoder and adds their scores inside the standard MaxSim.
The dense embedding is written into the list of token vectors as three slots, and a one-dimensional tag keeps slots and tokens from matching each other, so one model, one index and one search replace a hybrid pipeline.
The cross-encoder teacher is distilled into the fused score, the score that ranks documents at inference.

With 179M parameters, \sysv{} reaches 59.79 nDCG@10 on BEIR, 2.57 points above LateOn and ahead of it on all 15 datasets; only models with at least 1.5B parameters score higher.
It stays ahead on the decontaminated benchmark and reaches 70.55 on NanoBEIR, 83.77 on LoTTE and a recall@2 of 91.4 on LIMIT small.
Neither the slots nor the tag depends on the backbone or the language, so the construction carries over to multilingual and larger encoders.

\section*{Acknowledgments}
The authors would like to acknowledge the financial support provided by the Austrian Research Agency (FFG) for the project “AI Enabled Sustainability Jurisdiction Demonstrator” (project No. 915229). The computational results presented in this work have been achieved using the MUSICA cluster, part of the Austrian Scientific Computing (ASC) infrastructure.

\section*{Limitations}

The model is English only, and as a multi-vector model it needs a larger index than a dense model.
Like the other models trained on this collection, it is fine-tuned on the training splits of five BEIR datasets, and its auxiliary pairs come from the same sites as four more.
Cost is measured on one dataset, where search in PLAID is slower than for LateOn.

\bibliography{references}

\begin{thebibliography}{46}
\providecommand{\natexlab}[1]{#1}

\bibitem[{Abdallah et~al.(2026{\natexlab{a}})Abdallah, Ali, and Jatowt}]{abdallah2026smallreason}
Abdelrahman Abdallah, Mohammed Ali, and Adam Jatowt. 2026{\natexlab{a}}.
\newblock Smallreason-colbert: An ultra-small late-interaction retriever for reasoning intensive retrieval.
\newblock \emph{arXiv preprint arXiv:2609.29652}.

\bibitem[{Abdallah et~al.(2026{\natexlab{b}})Abdallah, Piryani, Mozafari, Herzinger, Holdcroft, and Jatowt}]{abdallah2026rankify}
Abdelrahman Abdallah, Bhawna Piryani, Jamshid Mozafari, Andreas Herzinger, Jamie Holdcroft, and Adam Jatowt. 2026{\natexlab{b}}.
\newblock Rankify: A comprehensive python toolkit for retrieval, re-ranking, and retrieval-augmented generation.
\newblock In \emph{Proceedings of the 64th Annual Meeting of the Association for Computational Linguistics (Volume 3: System Demonstrations)}, pages 208--219.

\bibitem[{Ali et~al.(2026)Ali, Abdallah, and Jatowt}]{ali2026excise}
Mohammed Ali, Abdelrahman Abdallah, and Adam Jatowt. 2026.
\newblock Excise: Query-side exclusion for late-interaction retrieval.
\newblock \emph{arXiv preprint arXiv:2608.05497}.

\bibitem[{Bajaj et~al.(2016)Bajaj, Campos, Craswell, Deng, Gao, Liu, Majumder, McNamara, Mitra, Nguyen, Rosenberg, Song, Stoica, Tiwary, and Wang}]{msmarco}
Payal Bajaj, Daniel Campos, Nick Craswell, Li~Deng, Jianfeng Gao, Xiaodong Liu, Rangan Majumder, Andrew McNamara, Bhaskar Mitra, Tri Nguyen, Mir Rosenberg, Xia Song, Alina Stoica, Saurabh Tiwary, and Tong Wang. 2016.
\newblock {MS} {MARCO}: A human generated {MAchine} reading {COmprehension} dataset.
\newblock \emph{arXiv preprint arXiv:1611.09268}.

\bibitem[{Chaffin and Sourty(2025)}]{pylate}
Antoine Chaffin and Rapha{\"e}l Sourty. 2025.
\newblock \href {https://doi.org/10.1145/3746252.3761608} {{PyLate}: Flexible training and retrieval for late interaction models}.
\newblock In \emph{Proceedings of the 34th ACM International Conference on Information and Knowledge Management (CIKM)}, pages 6334--6339.

\bibitem[{Chen et~al.(2024)Chen, Xiao, Zhang, Luo, Lian, and Liu}]{bgem3}
Jianlv Chen, Shitao Xiao, Peitian Zhang, Kun Luo, Defu Lian, and Zheng Liu. 2024.
\newblock M3-embedding: Multi-linguality, multi-functionality, multi-granularity text embeddings through self-knowledge distillation.
\newblock \emph{arXiv preprint arXiv:2402.03216}.

\bibitem[{Clavié(2024)}]{jacolbert}
Benjamin Clavié. 2024.
\newblock {JaColBERTv2}.5: Optimising multi-vector retrievers to create state-of-the-art japanese retrievers with constrained resources.
\newblock \emph{arXiv preprint arXiv:2407.20750}.

\bibitem[{Cormack et~al.(2009)Cormack, Clarke, and B{\"u}ttcher}]{rrf}
Gordon~V. Cormack, Charles L.~A. Clarke, and Stefan B{\"u}ttcher. 2009.
\newblock Reciprocal rank fusion outperforms {C}ondorcet and individual rank learning methods.
\newblock In \emph{Proceedings of the 32nd International ACM SIGIR Conference on Research and Development in Information Retrieval}, pages 758--759.

\bibitem[{de~Souza P.~Moreira et~al.(2024)de~Souza P.~Moreira, Osmulski, Xu, Ak, Schifferer, and Oldridge}]{nvretriever}
Gabriel de~Souza P.~Moreira, Radek Osmulski, Mengyao Xu, Ronay Ak, Benedikt Schifferer, and Even Oldridge. 2024.
\newblock {NV-Retriever}: Improving text embedding models with effective hard-negative mining.
\newblock \emph{arXiv preprint arXiv:2407.15831}.

\bibitem[{Formal et~al.(2021)Formal, Piwowarski, and Clinchant}]{splade}
Thibault Formal, Benjamin Piwowarski, and Stéphane Clinchant. 2021.
\newblock {SPLADE}: Sparse lexical and expansion model for first stage ranking.
\newblock In \emph{Proceedings of the 44th International ACM SIGIR Conference on Research and Development in Information Retrieval}.

\bibitem[{Gao et~al.(2021)Gao, Dai, and Callan}]{coil}
Luyu Gao, Zhuyun Dai, and Jamie Callan. 2021.
\newblock {COIL}: Revisit exact lexical match in information retrieval with contextualized inverted list.
\newblock In \emph{Proceedings of the 2021 Conference of the North American Chapter of the Association for Computational Linguistics (NAACL)}.

\bibitem[{Günther et~al.(2025)Günther, Sturua, Akram, Mohr, Ungureanu, Wang, Eslami, Martens, Werk, Wang, and Xiao}]{jinav4}
Michael Günther, Saba Sturua, Mohammad~Kalim Akram, Isabelle Mohr, Andrei Ungureanu, Bo~Wang, Sedigheh Eslami, Scott Martens, Maximilian Werk, Nan Wang, and Han Xiao. 2025.
\newblock jina-embeddings-v4: Universal embeddings for multimodal multilingual retrieval.
\newblock \emph{arXiv preprint arXiv:2506.18902}.

\bibitem[{Hinton et~al.(2015)Hinton, Vinyals, and Dean}]{hinton}
Geoffrey Hinton, Oriol Vinyals, and Jeff Dean. 2015.
\newblock Distilling the knowledge in a neural network.
\newblock \emph{arXiv preprint arXiv:1503.02531}.

\bibitem[{Hofstätter et~al.(2020)Hofstätter, Althammer, Schröder, Sertkan, and Hanbury}]{crossarch}
Sebastian Hofstätter, Sophia Althammer, Michael Schröder, Mete Sertkan, and Allan Hanbury. 2020.
\newblock Improving efficient neural ranking models with cross-architecture knowledge distillation.
\newblock \emph{arXiv preprint arXiv:2010.02666}.

\bibitem[{Hofstätter et~al.(2022)Hofstätter, Khattab, Althammer, Sertkan, and Hanbury}]{colberter}
Sebastian Hofstätter, Omar Khattab, Sophia Althammer, Mete Sertkan, and Allan Hanbury. 2022.
\newblock Introducing neural bag of whole-words with {ColBERTer}: Contextualized late interactions using enhanced reduction.
\newblock In \emph{Proceedings of the 31st ACM International Conference on Information and Knowledge Management (CIKM)}.

\bibitem[{Izacard et~al.(2022)Izacard, Caron, Hosseini, Riedel, Bojanowski, Joulin, and Grave}]{contriever}
Gautier Izacard, Mathilde Caron, Lucas Hosseini, Sebastian Riedel, Piotr Bojanowski, Armand Joulin, and Edouard Grave. 2022.
\newblock Unsupervised dense information retrieval with contrastive learning.
\newblock \emph{Transactions on Machine Learning Research}.

\bibitem[{Jaegle et~al.(2021)Jaegle, Gimeno, Brock, Zisserman, Vinyals, and Carreira}]{perceiver}
Andrew Jaegle, Felix Gimeno, Andrew Brock, Andrew Zisserman, Oriol Vinyals, and Joao Carreira. 2021.
\newblock Perceiver: General perception with iterative attention.
\newblock In \emph{Proceedings of the 38th International Conference on Machine Learning (ICML)}.

\bibitem[{Jha et~al.(2024)Jha, Wang, Günther, Mastrapas, Sturua, Mohr, Koukounas, Akram, Wang, and Xiao}]{jinacolbertv2}
Rohan Jha, Bo~Wang, Michael Günther, Georgios Mastrapas, Saba Sturua, Isabelle Mohr, Andreas Koukounas, Mohammad~Kalim Akram, Nan Wang, and Han Xiao. 2024.
\newblock Jina-{ColBERT-v2}: A general-purpose multilingual late interaction retriever.
\newblock \emph{arXiv preprint arXiv:2408.16672}.

\bibitem[{Karpukhin et~al.(2020)Karpukhin, O{\u{g}}uz, Min, Lewis, Wu, Edunov, Chen, and tau Yih}]{dpr}
Vladimir Karpukhin, Barlas O{\u{g}}uz, Sewon Min, Patrick Lewis, Ledell Wu, Sergey Edunov, Danqi Chen, and Wen tau Yih. 2020.
\newblock Dense passage retrieval for open-domain question answering.
\newblock In \emph{Proceedings of the 2020 Conference on Empirical Methods in Natural Language Processing (EMNLP)}.

\bibitem[{Khattab and Zaharia(2020)}]{colbert}
Omar Khattab and Matei Zaharia. 2020.
\newblock {ColBERT}: Efficient and effective passage search via contextualized late interaction over {BERT}.
\newblock In \emph{Proceedings of the 43rd International ACM SIGIR Conference on Research and Development in Information Retrieval}.

\bibitem[{Lee et~al.(2025)Lee, Roy, Xu, Raiman, Shoeybi, Catanzaro, and Ping}]{nvembed}
Chankyu Lee, Rajarshi Roy, Mengyao Xu, Jonathan Raiman, Mohammad Shoeybi, Bryan Catanzaro, and Wei Ping. 2025.
\newblock {NV-Embed}: Improved techniques for training {LLMs} as generalist embedding models.
\newblock In \emph{International Conference on Learning Representations (ICLR)}.

\bibitem[{Lee et~al.(2023)Lee, Dai, Duddu, Lei, Naim, Chang, and Zhao}]{xtr}
Jinhyuk Lee, Zhuyun Dai, Sai Meher~Karthik Duddu, Tao Lei, Iftekhar Naim, Ming-Wei Chang, and Vincent~Y. Zhao. 2023.
\newblock Rethinking the role of token retrieval in multi-vector retrieval.
\newblock In \emph{Advances in Neural Information Processing Systems}.

\bibitem[{Li et~al.(2023)Li, Zhang, Zhang, Long, Xie, and Zhang}]{gte}
Zehan Li, Xin Zhang, Yanzhao Zhang, Dingkun Long, Pengjun Xie, and Meishan Zhang. 2023.
\newblock Towards general text embeddings with multi-stage contrastive learning.
\newblock \emph{arXiv preprint arXiv:2308.03281}.

\bibitem[{{LightOn}(2025)}]{gtemoderncolbert}
{LightOn}. 2025.
\newblock {GTE-ModernColBERT-v1}.
\newblock Model card, \url{https://huggingface.co/lightonai/GTE-ModernColBERT-v1}.

\bibitem[{{LightOn}(2026)}]{colbertzero}
{LightOn}. 2026.
\newblock {ColBERT-Zero}.
\newblock Model card, \url{https://huggingface.co/lightonai/ColBERT-Zero}.

\bibitem[{Luan et~al.(2021)Luan, Eisenstein, Toutanova, and Collins}]{mebert}
Yi~Luan, Jacob Eisenstein, Kristina Toutanova, and Michael Collins. 2021.
\newblock Sparse, dense, and attentional representations for text retrieval.
\newblock \emph{Transactions of the Association for Computational Linguistics}.

\bibitem[{Merrick et~al.(2024)Merrick, Xu, Nuti, and Campos}]{arctic}
Luke Merrick, Danmei Xu, Gaurav Nuti, and Daniel Campos. 2024.
\newblock Arctic-embed: Scalable, efficient, and accurate text embedding models.
\newblock \emph{arXiv preprint arXiv:2405.05374}.

\bibitem[{{Mixedbread}(2025)}]{mxbairerank}
{Mixedbread}. 2025.
\newblock mxbai-rerank-large-v2.
\newblock Model card, \url{https://huggingface.co/mixedbread-ai/mxbai-rerank-large-v2}.

\bibitem[{{MTEB contributors}(2026)}]{mtebresults}
{MTEB contributors}. 2026.
\newblock {MTEB} results repository.
\newblock \url{https://github.com/embeddings-benchmark/results}.

\bibitem[{Muennighoff et~al.(2023)Muennighoff, Tazi, Magne, and Reimers}]{mteb}
Niklas Muennighoff, Nouamane Tazi, Loïc Magne, and Nils Reimers. 2023.
\newblock {MTEB}: Massive text embedding benchmark.
\newblock In \emph{Proceedings of the 17th Conference of the European Chapter of the Association for Computational Linguistics (EACL)}.

\bibitem[{Nussbaum et~al.(2024)Nussbaum, Morris, Duderstadt, and Mulyar}]{nomic}
Zach Nussbaum, John~X. Morris, Brandon Duderstadt, and Andriy Mulyar. 2024.
\newblock Nomic embed: Training a reproducible long context text embedder.
\newblock \emph{arXiv preprint arXiv:2402.01613}.

\bibitem[{Reimers and Gurevych(2019)}]{sbert}
Nils Reimers and Iryna Gurevych. 2019.
\newblock Sentence-{BERT}: Sentence embeddings using siamese {BERT-Networks}.
\newblock In \emph{Proceedings of the 2019 Conference on Empirical Methods in Natural Language Processing (EMNLP-IJCNLP)}.

\bibitem[{Ren et~al.(2021)Ren, Qu, Liu, Zhao, She, Wu, Wang, and Wen}]{rocketqav2}
Ruiyang Ren, Yingqi Qu, Jing Liu, Wayne~Xin Zhao, Qiaoqiao She, Hua Wu, Haifeng Wang, and Ji-Rong Wen. 2021.
\newblock {RocketQAv2}: A joint training method for dense passage retrieval and passage re-ranking.
\newblock In \emph{Proceedings of the 2021 Conference on Empirical Methods in Natural Language Processing (EMNLP)}.

\bibitem[{Santhanam et~al.(2022{\natexlab{a}})Santhanam, Khattab, Potts, and Zaharia}]{plaid}
Keshav Santhanam, Omar Khattab, Christopher Potts, and Matei Zaharia. 2022{\natexlab{a}}.
\newblock {PLAID}: An efficient engine for late interaction retrieval.
\newblock In \emph{Proceedings of the 31st ACM International Conference on Information and Knowledge Management (CIKM)}.

\bibitem[{Santhanam et~al.(2022{\natexlab{b}})Santhanam, Khattab, Saad-Falcon, Potts, and Zaharia}]{colbertv2}
Keshav Santhanam, Omar Khattab, Jon Saad-Falcon, Christopher Potts, and Matei Zaharia. 2022{\natexlab{b}}.
\newblock {ColBERTv2}: Effective and efficient retrieval via lightweight late interaction.
\newblock In \emph{Proceedings of the 2022 Conference of the North American Chapter of the Association for Computational Linguistics (NAACL)}.

\bibitem[{Shen et~al.(2023)Shen, Geng, Tao, Xu, Long, Zhang, and Jiang}]{unifier}
Tao Shen, Xiubo Geng, Chongyang Tao, Can Xu, Guodong Long, Kai Zhang, and Daxin Jiang. 2023.
\newblock {UnifieR}: A unified retriever for large-scale retrieval.
\newblock In \emph{Proceedings of the 29th ACM SIGKDD Conference on Knowledge Discovery and Data Mining (KDD)}.

\bibitem[{Sourty et~al.(2026)Sourty, Chaffin, Weller, Demoura, and Chatelain}]{lateon}
Raphael Sourty, Antoine Chaffin, Orion Weller, Paulo Demoura, and Amelie Chatelain. 2026.
\newblock {DenseOn} with the {LateOn}: Open state-of-the-art single and multi-vector models.
\newblock \url{https://huggingface.co/blog/lightonai/denseon-lateon}.

\bibitem[{Thakur et~al.(2021)Thakur, Reimers, Rücklé, Srivastava, and Gurevych}]{beir}
Nandan Thakur, Nils Reimers, Andreas Rücklé, Abhishek Srivastava, and Iryna Gurevych. 2021.
\newblock {BEIR}: A heterogenous benchmark for zero-shot evaluation of information retrieval models.
\newblock In \emph{Advances in Neural Information Processing Systems, Datasets and Benchmarks Track}.

\bibitem[{Wang et~al.(2022)Wang, Yang, Huang, Jiao, Yang, Jiang, Majumder, and Wei}]{e5}
Liang Wang, Nan Yang, Xiaolong Huang, Binxing Jiao, Linjun Yang, Daxin Jiang, Rangan Majumder, and Furu Wei. 2022.
\newblock Text embeddings by weakly-supervised contrastive pre-training.
\newblock \emph{arXiv preprint arXiv:2212.03533}.

\bibitem[{Warner et~al.(2025)Warner, Chaffin, Clavié, Weller, Hallström, Taghadouini, Gallagher, Biswas, Ladhak, Aarsen, Cooper, Adams, Howard, and Poli}]{modernbert}
Benjamin Warner, Antoine Chaffin, Benjamin Clavié, Orion Weller, Oskar Hallström, Said Taghadouini, Alexis Gallagher, Raja Biswas, Faisal Ladhak, Tom Aarsen, Nathan Cooper, Griffin Adams, Jeremy Howard, and Iacopo Poli. 2025.
\newblock Smarter, better, faster, longer: A modern bidirectional encoder for fast, memory efficient, and long context finetuning and inference.
\newblock In \emph{Proceedings of the 63rd Annual Meeting of the Association for Computational Linguistics (ACL)}.

\bibitem[{Weller et~al.(2025)Weller, Boratko, Naim, and Lee}]{limit}
Orion Weller, Michael Boratko, Iftekhar Naim, and Jinhyuk Lee. 2025.
\newblock On the theoretical limitations of embedding-based retrieval.
\newblock \emph{arXiv preprint arXiv:2508.21038}.

\bibitem[{Wortsman et~al.(2022)Wortsman, Ilharco, Gadre, Roelofs, Gontijo-Lopes, Morcos, Namkoong, Farhadi, Carmon, Kornblith, and Schmidt}]{soups}
Mitchell Wortsman, Gabriel Ilharco, Samir~Yitzhak Gadre, Rebecca Roelofs, Raphael Gontijo-Lopes, Ari~S. Morcos, Hongseok Namkoong, Ali Farhadi, Yair Carmon, Simon Kornblith, and Ludwig Schmidt. 2022.
\newblock Model soups: averaging weights of multiple fine-tuned models improves accuracy without increasing inference time.
\newblock In \emph{Proceedings of the 39th International Conference on Machine Learning (ICML)}.

\bibitem[{Xiao et~al.(2024)Xiao, Liu, Zhang, Muennighoff, Lian, and Nie}]{bge}
Shitao Xiao, Zheng Liu, Peitian Zhang, Niklas Muennighoff, Defu Lian, and Jian-Yun Nie. 2024.
\newblock C-pack: Packed resources for general chinese embeddings.
\newblock In \emph{Proceedings of the 47th International ACM SIGIR Conference on Research and Development in Information Retrieval}.

\bibitem[{Xiong et~al.(2021)Xiong, Xiong, Li, Tang, Liu, Bennett, Ahmed, and Overwijk}]{ance}
Lee Xiong, Chenyan Xiong, Ye~Li, Kwok-Fung Tang, Jialin Liu, Paul Bennett, Junaid Ahmed, and Arnold Overwijk. 2021.
\newblock Approximate nearest neighbor negative contrastive learning for dense text retrieval.
\newblock In \emph{International Conference on Learning Representations (ICLR)}.

\bibitem[{{Zeta Alpha}(2024)}]{nanobeir}
{Zeta Alpha}. 2024.
\newblock {NanoBEIR}: a collection of smaller versions of {BEIR} datasets.
\newblock \url{https://huggingface.co/collections/zeta-alpha-ai/nanobeir}.

\bibitem[{Zhang et~al.(2024)Zhang, Li, Zeng, and Wang}]{stella}
Dun Zhang, Jiacheng Li, Ziyang Zeng, and Fulong Wang. 2024.
\newblock Jasper and stella: distillation of {SOTA} embedding models.
\newblock \emph{arXiv preprint arXiv:2412.19048}.

\end{thebibliography}

\appendix
\section{Training and Evaluation Details}
\label{app:details}

Table~\ref{tab:settings} lists the settings of the released model.

\begin{table}[h]
\centering
\small
\resizebox{\columnwidth}{!}{%
\begin{tabular}{ll}
\toprule
\multicolumn{2}{l}{\textit{Model}} \\
Backbone & ModernBERT-base, 22 layers \\
Shared layers / layers per top & 18 / 4 \\
Parameters & 179.0M \\
\quad backbone (trunk and token top) & 149.0M \\
\quad dense top & 20.1M \\
\quad pooling and projection & 9.9M \\
Latent attention & 512 latents, 8 heads \\
Vector dimension & 128 (1 tag + 127) \\
Slots & 3 \\
$\beta$ / $\delta$ / $\lambda$ & 0.9 / 0.225 / 0.75 \\
Prompts & \texttt{query:} / \texttt{document:} \\
\midrule
\multicolumn{2}{l}{\textit{Pre-training (stage 1)}} \\
Steps / batch & 33{,}000 / 16{,}384 pairs \\
Data & 331M pairs \\
Temperature & 0.02 \\
\midrule
\multicolumn{2}{l}{\textit{Last stage}} \\
Steps / batch & 16{,}597 / 128 queries \\
Learning rate / warm-up & $7\times10^{-5}$ / 5\% \\
Hard negatives per query & 14 \\
Query / document length & 256 / 512 tokens \\
Temperature (contrastive, student) & 0.02 \\
Temperature (teacher) & 0.1 \\
Anchor weight $\gamma$ & 10 \\
Hardware / time & 4 H100 / 4.3 hours \\
Pre-training time & about 75 hours on 4 H100 \\
\midrule
\multicolumn{2}{l}{\textit{Inference}} \\
Query / document length & 256 / 2048 tokens \\
Query tokens in token matching & first 64 \\
Precision & bfloat16 \\
\bottomrule
\end{tabular}}
\caption{Settings of \sysv{}.}
\label{tab:settings}
\end{table}

\paragraph{Model size.} Of the 179.0M parameters, 149.0M are those of ModernBERT-base, which form the trunk and the token top. The dense top, a second copy of the last four layers, adds 20.1M, and the latent-attention pooling with the projection adds 9.9M. A stored vector has 128 dimensions, as in LateOn and ColBERT-Zero: one dimension holds the tag and 127 hold the token or slot vector, so an index entry has the same size as theirs.

\paragraph{Training.} Pre-training is the expensive part: 33{,}000 steps with a batch of 16{,}384 pairs take about 75 hours on four H100 GPUs. The last stage, the final epoch of the two-top model, has 16{,}597 steps with a batch of 128 queries, each with its positive and 14 hard negatives, and takes 4.3 hours on the same hardware. It reads queries of up to 256 and documents of up to 512 tokens. The contrastive losses and the student distributions of the KL terms use a temperature of 0.02, the teacher distribution a temperature of 0.1.

\paragraph{Inference.} The model reads documents of up to 2048 tokens, four times the training length, and queries of up to 256 tokens, of which the first 64 take part in the token-level matching. The encoder runs in bfloat16.

\paragraph{Overlap with BEIR.} Rows whose query equals a BEIR test or development query were removed from the training data: 37{,}391 rows of the pre-training sample and 228 rows of the fine-tuning collection. For the mined auxiliary pairs, a pair was removed if its query or its document equals a BEIR query.

\paragraph{Teachers.} The cross-encoder teacher is mxbai-rerank-large-v2 \citep{mxbairerank}. Three of the four dense fine-tunes that are averaged into the dense model of stage 2 were trained with an additional ranking loss to a fine-tune of stella\_en\_400M\_v5 \citep{stella}.

\section{Additional Results}
\label{app:results}

This appendix gives the full BEIR and NanoBEIR tables, the tables behind the LIMIT and cost results of \S\ref{sec:results}, the LoTTE result per domain and the effect of the document length.

\begin{table*}[t]
\centering
\resizebox{\textwidth}{!}{%
\begin{tabular}{lrccccccccccccccc|c}
\toprule
Model & Size & Arg. & CQA & C-FEV & DBP & FEV & FiQA & HQA & MSM & NFC & NQ & Quora & SCID & SciF & COV & Tou. & Avg. \\
\midrule
\multicolumn{18}{l}{\textit{Single-vector models, at most 250M parameters}} \\
bge-base-en & 109M & 61.49 & 41.35 & 31.61 & 42.01 & 88.35 & 40.76 & 72.35 & 41.26 & 35.54 & 53.98 & 88.78 & 21.30 & 73.26 & 76.45 & 26.54 & 53.00 \\
modernbert-embed-base & 149M & 48.96 & 42.08 & 35.67 & 41.49 & 87.35 & 40.59 & 67.11 & 41.47 & 33.40 & 62.15 & 88.85 & 18.59 & 69.63 & \underline{84.15} & 31.91 & 52.89 \\
granite-embedding-english-r2 & 149M & 59.21 & \textbf{50.00} & 35.82 & 39.60 & 88.04 & 46.32 & 67.36 & 32.14 & 37.49 & 58.22 & 87.84 & \textbf{24.95} & 75.80 & 70.56 & 22.90 & 53.08 \\
gte-base-en-v1.5 & 137M & 63.49 & 39.52 & 40.36 & 39.90 & \textbf{94.81} & 48.65 & 67.75 & 42.62 & 35.88 & 52.96 & 88.42 & 21.92 & \underline{76.77} & 73.13 & 25.22 & 54.09 \\
cadet-embed-base-v1 & 109M & 56.16 & 40.80 & 36.69 & 45.26 & 89.33 & 40.89 & 73.76 & 42.56 & 37.71 & 59.32 & 87.59 & 19.09 & 74.68 & 80.55 & 30.42 & 54.32 \\
snowflake-arctic-embed-m & 109M & 56.44 & 43.81 & 39.37 & 44.73 & 89.02 & 42.40 & 73.65 & 41.77 & 36.77 & 62.43 & 87.42 & 21.10 & 73.55 & 79.65 & 31.47 & 54.91 \\
gte-modernbert-base & 149M & \textbf{74.56} & 42.64 & \textbf{45.90} & 41.39 & \underline{93.97} & 49.54 & 70.39 & 39.93 & 34.32 & 56.10 & 88.56 & 20.44 & 76.41 & 75.75 & 17.97 & 55.19 \\
jina-embeddings-v5-text-nano & 212M & \underline{65.70} & 44.65 & 39.60 & 45.26 & 89.51 & 47.85 & 69.07 & 41.64 & 38.69 & 63.38 & 88.87 & 22.60 & 75.78 & 77.60 & 30.70 & 56.06 \\
DenseOn & 149M & 54.65 & 46.89 & 37.49 & 44.65 & 90.69 & \textbf{53.86} & 74.51 & 43.58 & \textbf{39.03} & 59.25 & 89.31 & 22.35 & 75.95 & 82.33 & 28.43 & 56.20 \\
\midrule
\multicolumn{18}{l}{\textit{Late-interaction models, at most 250M parameters}} \\
GTE-ModernColBERT-v1 & 149M & 47.52 & 41.08 & 31.33 & \underline{47.56} & 87.67 & 45.25 & 77.48 & 45.60 & 37.83 & 61.62 & 86.71 & 19.22 & 76.33 & \textbf{84.84} & 31.25 & 54.75 \\
ColBERT-Zero & 149M & 52.81 & 41.41 & 35.90 & 47.43 & 90.52 & 42.50 & 79.45 & \underline{45.95} & 37.21 & 61.82 & 85.19 & 19.84 & 76.33 & 78.27 & \underline{36.24} & 55.39 \\
LateOn & 149M & 50.52 & 47.36 & 39.67 & 45.99 & 92.02 & 53.12 & \underline{79.98} & 45.67 & 37.79 & \underline{63.91} & \underline{89.66} & 21.90 & 76.61 & 83.60 & 30.52 & \underline{57.22} \\
\textbf{UIBK-Embedding-v1 (ours)} & 179M & 56.56 & \underline{48.55} & \underline{44.89} & \textbf{49.22} & 93.28 & \underline{53.73} & \textbf{81.50} & \textbf{46.82} & \underline{38.79} & \textbf{67.80} & \textbf{89.92} & \underline{23.60} & \textbf{78.57} & 83.82 & \textbf{39.78} & \textbf{59.79} \\
\midrule
\multicolumn{18}{l}{\textit{Larger models, for reference}} \\
snowflake-arctic-embed-l & 334M & 59.09 & 46.97 & 39.33 & 45.97 & 88.22 & 44.71 & 75.18 & 41.68 & 37.65 & 63.11 & 87.41 & 21.36 & 73.82 & 80.72 & 34.52 & 55.98 \\
Qwen3-Embedding-0.6B & 596M & 70.97 & 46.03 & 42.11 & 39.48 & 88.15 & 46.61 & 65.74 & 37.99 & 36.71 & 53.46 & 87.78 & 24.41 & 69.72 & 90.52 & 33.18 & 55.52 \\
jina-embeddings-v5-text-small & 596M & 65.07 & 46.72 & 41.50 & 44.38 & 89.99 & 49.63 & 69.76 & 42.09 & 39.81 & 64.04 & 89.09 & 23.04 & 76.53 & 78.49 & 29.89 & 56.67 \\
mDenseOn & 307M & 55.31 & 45.70 & 41.51 & 43.61 & 91.46 & 51.91 & 75.28 & 43.08 & 37.34 & 63.49 & 88.91 & 21.51 & 74.88 & 81.63 & 34.91 & 56.70 \\
mLateOn & 307M & 59.81 & 44.67 & 39.84 & 48.39 & 93.89 & 49.93 & 82.24 & 44.90 & 37.63 & 65.09 & 88.39 & 20.26 & 76.35 & 82.12 & 29.94 & 57.56 \\
Qwen3-Embedding-4B & 4.0B & 75.64 & 50.28 & 47.43 & 48.23 & 91.60 & 62.65 & 74.72 & 42.72 & 41.10 & 63.13 & 88.06 & 31.44 & 78.33 & 92.92 & 35.41 & 61.58 \\
Qwen3-Embedding-8B & 7.6B & 76.85 & 52.94 & 47.36 & 49.68 & 91.88 & 64.57 & 76.78 & 43.60 & 41.45 & 65.25 & 88.90 & 32.74 & 78.46 & 94.99 & 35.93 & 62.76 \\
\bottomrule
\end{tabular}

}
\caption{BEIR, nDCG@10 on every dataset: Table~\ref{tab:beir} with the larger models. All numbers are from the official MTEB results repository. \textbf{Bold} and \underline{underlined}: best and second best among the models with at most 250M parameters.}
\label{tab:beir-full}
\end{table*}

\begin{table*}[t]
\centering
\resizebox{0.9\textwidth}{!}{%
\begin{tabular}{lrccccccccccccc|c}
\toprule
Model & Size & Arg. & C-FEV & DBP & FEV & FiQA & HQA & MSM & NFC & NQ & Quora & SCID & SciF & Tou. & Avg. \\
\midrule
\multicolumn{16}{l}{\textit{At most 250M parameters}} \\
BM25 & -- & 48.04 & 31.77 & 61.38 & 80.94 & 43.74 & \underline{82.77} & 53.30 & 32.50 & 50.07 & 86.28 & 33.51 & 70.99 & \textbf{69.53} & 57.29 \\
multilingual-e5-small & 118M & 44.54 & 30.64 & 60.54 & 83.49 & 43.53 & 81.01 & \underline{62.09} & 28.82 & 68.62 & \textbf{97.28} & 34.38 & 72.46 & 49.54 & 58.23 \\
F2LLM-v2-80M & 80M & 53.45 & 39.19 & 56.99 & \underline{92.89} & 44.89 & 72.66 & 56.91 & 32.63 & 54.59 & 95.86 & 38.54 & 66.89 & 53.84 & 58.41 \\
F2LLM-v2-160M & 159M & 57.07 & \underline{41.45} & 57.75 & 92.13 & 48.74 & 73.56 & 57.59 & 33.50 & 60.33 & 95.57 & \underline{40.59} & 67.76 & 53.11 & 59.93 \\
bekko-embedding-v1-a8m & 106M & 54.72 & 31.18 & 60.34 & 87.38 & 51.48 & 78.52 & 58.50 & 34.13 & 63.17 & 93.51 & 39.53 & 76.67 & 55.92 & 60.39 \\
bekko-embedding-v1-a25m & 123M & \textbf{60.73} & 26.06 & \underline{63.17} & 85.19 & \underline{53.39} & 80.91 & 61.46 & \underline{34.46} & \underline{69.10} & 94.24 & 39.46 & \underline{78.17} & 57.04 & \underline{61.80} \\
\textbf{UIBK-Embedding-v1 (ours)} & 179M & \underline{59.84} & \textbf{51.65} & \textbf{69.20} & \textbf{96.57} & \textbf{64.34} & \textbf{92.72} & \textbf{68.99} & \textbf{39.22} & \textbf{82.14} & \underline{96.99} & \textbf{45.40} & \textbf{83.93} & \underline{66.14} & \textbf{70.55} \\
\midrule
\multicolumn{16}{l}{\textit{Larger models, for reference}} \\
F2LLM-v2-330M & 334M & 60.11 & 46.17 & 59.95 & 94.84 & 56.12 & 76.58 & 61.79 & 34.38 & 65.89 & 95.47 & 41.45 & 77.75 & 54.75 & 63.48 \\
F2LLM-v2-0.6B & 596M & 58.16 & 48.31 & 61.76 & 95.28 & 58.23 & 78.97 & 69.12 & 37.77 & 70.45 & 96.09 & 42.15 & 75.61 & 51.08 & 64.84 \\
F2LLM-v2-1.7B & 1.7B & 55.46 & 45.09 & 64.57 & 93.52 & 66.78 & 81.99 & 66.64 & 36.22 & 73.07 & 96.70 & 44.74 & 81.53 & 51.69 & 66.00 \\
inf-retriever-v1-1.5b & 1.5B & 77.39 & 46.67 & 69.86 & 96.08 & 64.41 & 86.88 & 67.71 & 40.55 & 73.44 & 97.15 & 49.46 & 84.36 & 52.18 & 69.70 \\
F2LLM-v2-4B & 4.0B & 62.20 & 47.33 & 67.57 & 96.78 & 66.69 & 82.58 & 72.64 & 38.93 & 76.65 & 97.28 & 48.49 & 84.33 & 54.24 & 68.90 \\
F2LLM-v2-8B & 7.6B & 60.26 & 44.78 & 69.09 & 95.06 & 68.27 & 86.60 & 70.52 & 38.79 & 81.32 & 97.74 & 49.41 & 85.59 & 54.95 & 69.41 \\
F2LLM-v2-14B & 14.0B & 62.29 & 45.64 & 70.30 & 95.43 & 68.63 & 88.09 & 72.75 & 41.20 & 82.59 & 97.57 & 50.93 & 82.20 & 56.13 & 70.29 \\
inf-retriever-v1 & 7.1B & 73.21 & 43.25 & 73.45 & 97.59 & 69.72 & 90.95 & 70.06 & 47.10 & 78.31 & 97.22 & 53.33 & 86.32 & 56.03 & 72.04 \\
\bottomrule
\end{tabular}

}
\caption{NanoBEIR, nDCG@10 on every dataset: Table~\ref{tab:nanobeir} with the larger models. \textbf{Bold} and \underline{underlined} as in Table~\ref{tab:beir-full}.}
\label{tab:nanobeir-full}
\end{table*}

\paragraph{Full BEIR and NanoBEIR tables.} Tables~\ref{tab:beir-full} and~\ref{tab:nanobeir-full} repeat Tables~\ref{tab:beir} and~\ref{tab:nanobeir} with the larger models that have official results. On BEIR, the five models with 307M to 596M parameters reach between 55.52 and 57.56, and \sysv{} is ahead of each of them on at least 12 of the 15 datasets. Qwen3-Embedding-4B and 8B, with 22 and 42 times as many parameters, reach 61.58 and 62.76; \sysv{} is ahead of them on 8 and 7 datasets. On NanoBEIR, \sysv{} is above every model of the table except inf-retriever-v1 with 7.1B parameters, including F2LLM-v2-14B, which has 78 times as many parameters.

\paragraph{LIMIT.} LIMIT \citep{limit} asks for the two documents that are relevant to a query. Its queries and documents are simple by design; the task is hard because a model must be able to return many different pairs of documents. Table~\ref{tab:limit} gives recall@2 on LIMIT small, with the numbers that \citet{limit} report for other neural retrievers. The single-vector models, with up to 8B parameters, reach between 19.0 and 54.3. GTE-ModernColBERT, a late-interaction model with 149M parameters, reaches 83.5, and \sysv{} reaches 91.4, 7.9 points more with vectors of the same dimension.

\begin{table}[h]
\centering
\small
\begin{tabular}{lrc}
\toprule
Model & Size & Recall@2 \\
\midrule
\textbf{\sysv{} (ours)} & 179M & \textbf{91.4} \\
GTE-ModernColBERT & 149M & 83.5 \\
Promptriever (Llama3) & 8B & 54.3 \\
GritLM & 7B & 38.4 \\
Gemini Embed & -- & 33.7 \\
Qwen3 Embed & 8B & 19.0 \\
\bottomrule
\end{tabular}
\caption{LIMIT small (46 documents), recall@2 of neural retrievers. The numbers of the other models are from \citet{limit}; the lexical baseline BM25 reaches 97.8.}
\label{tab:limit}
\end{table}

\paragraph{LoTTE per domain.} LoTTE is scored by success@5, the share of queries with a relevant document among the first five results, and Table~\ref{tab:lotte} gives the mean over its five domains and two query types. Table~\ref{tab:lotte-domains} gives the result of \sysv{} for each of them. Success@5 is highest for Lifestyle (91.68 and 91.46) and Writing (89.64 and 89.60) and lowest for Science (71.15 and 69.86). The two query types are close in every domain, with means of 83.97 for search queries and 83.56 for forum queries.

\begin{table}[h]
\centering
\small
\begin{tabular}{lcc}
\toprule
Domain & Search & Forum \\
\midrule
Writing & 89.64 & 89.60 \\
Recreation & 85.82 & 87.36 \\
Science & 71.15 & 69.86 \\
Technology & 81.54 & 79.54 \\
Lifestyle & 91.68 & 91.46 \\
\midrule
Mean & 83.97 & 83.56 \\
\bottomrule
\end{tabular}
\caption{\sysv{} on the test split of LoTTE, success@5 per domain and query type.}
\label{tab:lotte-domains}
\end{table}

\paragraph{Cost.} Table~\ref{tab:cost} gives the measurements that \S\ref{sec:results} summarises. Every model encodes the 57{,}638 documents of FiQA on one H100 GPU, with queries cut at 32 and documents at 300 tokens, so the three encoders read the same text. \sysv{} writes 7.75M vectors and LateOn 7.70M, and the two PLAID indexes differ by 1\% (580 and 574~MiB). Encoding is 9\% slower than for LateOn, which reflects the four additional layers of the dense top. The last row is the hybrid of the two LightOn models: it encodes every document twice and keeps two indexes, 743~MiB together. Search times are means over the 648 test queries.

\begin{table}[h]
\centering
\small
\begin{tabular}{lrrr}
\toprule
Model & Docs/s & Index & Search \\
 & & (MiB) & (ms/query) \\
\midrule
\sysv{} & 899 & 580 & 7.8 \\
LateOn & 986 & 574 & 4.4 \\
DenseOn & 1{,}201 & 169 & 0.04 \\
LateOn + DenseOn & 541 & 743 & 4.4 \\
\bottomrule
\end{tabular}
\caption{Cost on FiQA (57{,}638 documents, 648 queries) on one H100 GPU, with queries of 32 and documents of 300 tokens for every model. Late-interaction models use a PLAID index with default settings; DenseOn uses a flat float32 index searched exhaustively. The last row adds the two models and does not include the fusion step.}
\label{tab:cost}
\end{table}

\paragraph{Document length.} The model is trained with documents of up to 512 tokens and reads longer ones at inference. A two-top model scores 57.61 on BEIR with 512 tokens, 58.14 with 2048 and 58.04 with 8192: the longer input gives 0.53 points, and the gain is complete at 2048 tokens, which is the released setting.

\section{Scoring and Bounds}
\label{app:bounds}

\begin{figure}[h]
\centering
\includegraphics[width=\columnwidth]{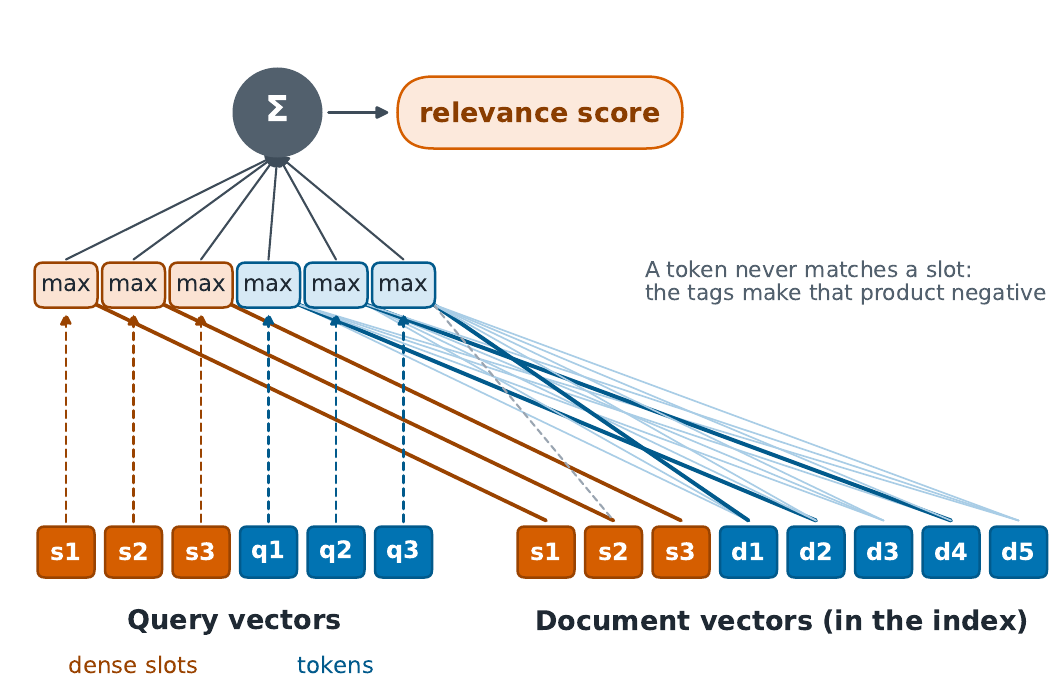}
\caption{Scoring is the standard MaxSim. Every query vector takes its best match among the document vectors. Because of the tag, a token only matches tokens (blue) and a slot only its own slot (orange), so the sum is a late-interaction score plus a dense score.}
\label{fig:scoring}
\end{figure}

\paragraph{Scoring.} Figure~\ref{fig:scoring} illustrates Eq.~\ref{eq:fused}. The query and the document are each one list that holds the three slots and the token vectors, and MaxSim takes, for every query vector, its best match in the document list. The tag gives every product of a token with a slot a negative offset, so a query token finds its best match among the document tokens and a query slot in the document slot with the same number. The sum over the query vectors is then the late-interaction score plus the weighted dense score.

\paragraph{Bounds as a function of the weight.} Table~\ref{tab:bounds} gives the worst-case quantities of \S\ref{sec:onelist} when the weight $\lambda$ of the dense score is changed through $\delta = \lambda\beta/3$ with $\beta = 0.9$.
A larger $\delta$ moves norm from the tag of a slot to its content, so the smallest slot--slot product, $1-2\delta$, falls and the largest slot--token product, $\sqrt{\beta\delta}-\sqrt{(1-\beta)(1-\delta)}$, rises.
Property~(i) holds as long as the first is above the second, which is the case up to $\lambda = 1$.
For property~(ii) the table gives the cosine that the best token match of a query token must exceed in the worst case.
The released value $\lambda = 0.75$ leaves a margin of 0.38 between the two products of property~(i) and asks for a best token cosine of only 0.08.
The fused score is flat for $\lambda$ between 0.75 and 1 (Figure~\ref{fig:ablations}, middle), the upper part of the range in which property~(i) holds.

\begin{table}[h]
\centering
\small
\resizebox{\columnwidth}{!}{%
\begin{tabular}{cccccc}
\toprule
$\lambda$ & $\delta$ & slot--slot & slot--token & (i) & (ii): cosine \\
 & & min. & max. & holds & above \\
\midrule
0.50 & 0.150 & 0.70 & 0.08 & yes & $-0.03$ \\
0.75 & 0.225 & 0.55 & 0.17 & yes & 0.08 \\
1.00 & 0.300 & 0.40 & 0.26 & yes & 0.17 \\
1.50 & 0.450 & 0.10 & 0.40 & no & 0.34 \\
2.50 & 0.750 & $-0.50$ & 0.66 & no & 0.63 \\
\bottomrule
\end{tabular}}
\caption{Worst-case bounds for the weights of Figure~\ref{fig:ablations} (middle). The released model uses $\lambda = 0.75$.}
\label{tab:bounds}
\end{table}

\section{Auxiliary Training Pairs}
\label{app:aux}

Table~\ref{tab:aux} lists the nine sources of the pairs mined from the pre-training collection (\S\ref{sec:data}).
The first six give 412k pairs with short queries, 10 to 12 words on average: duplicate questions from Quora and StackExchange, question--answer and title--body pairs from StackExchange, and citation and title--abstract pairs from S2ORC.
The last three give 152k pairs whose query is a whole paragraph, 92 to 173 words on average: a StackExchange question body with a related body, an S2ORC abstract with the abstract of a cited paper, and a Reddit post with a reply.
The supervised queries are short questions, so these pairs are where the dense head learns to read long queries.

\begin{table}[h]
\centering
\small
\resizebox{\columnwidth}{!}{%
\begin{tabular}{lrr}
\toprule
Source & Pairs & Query words \\
\midrule
Quora, duplicate questions & 35{,}206 & 10 \\
StackExchange, duplicate questions & 54{,}885 & 10 \\
StackExchange, question--answer & 91{,}973 & 10 \\
StackExchange, title--body & 98{,}831 & 10 \\
S2ORC, citation & 81{,}090 & 12 \\
S2ORC, title--abstract & 49{,}728 & 11 \\
\midrule
StackExchange, body--body & 55{,}545 & 92 \\
S2ORC, abstract--cited abstract & 56{,}956 & 173 \\
Reddit, post--reply & 39{,}024 & 158 \\
\bottomrule
\end{tabular}}
\caption{Auxiliary pairs: source, number of pairs and mean query length in words.}
\label{tab:aux}
\end{table}

\end{document}